\documentclass[twocolumn,showpacs,preprintnumbers,superscriptaddress,amsmath,amssymb,floatfix,prd]{revtex4}
\usepackage[colorlinks=true]{hyperref}

\newcommand{\comment}[1]{}

\newcommand{\lr}[1]{ \left( #1 \right) }
\newcommand{\lrs}[1]{ \left[ #1 \right] }

\newcommand{\tr}{ {\rm Tr} \: }

\newcommand{\sign}{ {\rm sign} \:  }
\newcommand{\sh}{ {\rm sh} \:  }
\newcommand{\ch}{ {\rm ch} \:  }
\renewcommand{\th}{ {\rm th} \:  }
\newcommand{\hodge}{{}^{*}}
\newcommand{\expa}[1]{ \exp{\left( #1 \right)} }

\begin{document}
\sloppy
\preprint{ITEP-LAT/2008-01}

\title{$Z_{2}$ electric strings and center vortices in $SU\lr{2}$ lattice gauge theory}
\author{M. I. Polikarpov}
\email{polykarp@itep.ru}
\affiliation{ITEP, B. Cheremushkinskaya str. 25, 117218 Moscow, Russia}
\author{P. V. Buividovich}
\email{buividovich@tut.by}
\affiliation{JIPNR, National Academy of Science, 220109 Belarus, Minsk, Acad. Krasin str. 99}
\affiliation{ITEP, B. Cheremushkinskaya str. 25, 117218 Moscow, Russia}
\date{January 1, 2008}
\begin{abstract}
We study the representations of $SU\lr{2}$ lattice gauge theory in terms of sums over the worldsheets of center vortices and $Z_{2}$ electric strings, i.e. surfaces which open on the Wilson loop. It is shown that in contrast to center vortices the density of electric $Z_{2}$ strings diverges in the continuum limit of the theory independently of the gauge fixing, however, their contribution to the Wilson loop yields physical string tension due to non-positivity of their statistical weight in the path integral, which is in turn related to the presence of $Z_{2}$ topological monopoles in the theory.
\end{abstract}
\maketitle

 It is often believed that Yang-Mills theory can be entirely reformulated in terms of string degrees of freedom, since the basic property of non-Abelian gauge theories -- quark confinement -- is explained by the emergence of confining string which stretches between test quark and antiquark. Even the simplest string models for Yang-Mills theory turn out to be very successful in reproducing the spectrum of bound states of the theory. One of the most successful recent developments is the AdS/QCD, the description of QCD bound states in terms of string theory on five-dimensional anti-de Sitter space or its modifications \cite{Polyakov:04:1, Son:06:1}. In this description the Wilson loop behaves as $W\lrs{C} \sim \expa{ - S_{5D}\lrs{C}}$, where $S_{5D}\lrs{C}$ is the area of the minimal surface in five-dimensional space spanned on the loop $C$. The loop $C$ is assumed to lie on the boundary of this five-dimensional space. However, up to now there is no exact representation of continuum Yang-Mills theory in terms of electric strings, i.e. the strings which open on Wilson loops. In fact, the only string one usually encounters in non-Abelian gauge theories is a chromoelectric string of finite thickness at finite lattice spacing, which is observed in numerical simulations as a cylindric region with higher energy density between two test colour charges \cite{Bali:95:1, Gubarev:07:1}.

 Recently a different type of strings has been discovered in lattice gauge theories, namely, $Z_{N}$ magnetic strings or center vortices. Although the existence of such strings in Yang-Mills theories was predicted a long time ago \cite{tHooft:78}, they have been actually observed and investigated in lattice simulations only during the last decade \cite{DelDebbio:98, Polikarpov:03:1}. The simulations has shown that center vortices are infinitely thin and have a finite density in the continuum limit. Moreover, lattice results suggest that center vortices are the effective degrees of freedom in the infrared domain of Yang-Mills theory \cite{DelDebbio:98, Polikarpov:03:1}, since removing center vortices from lattice configurations destroys all its characteristic infrared properties, such as confinement or spontaneous breaking of chiral symmetry. Effective action of center vortices and their geometric properties were extensively studied in \cite{Buividovich:07:3}. Detection of center vortices in lattice configurations of gauge fields is based on the separation of $SU\lr{N}$ link variables into $SU\lr{N}/Z_{N}$ variables and the variables which take values in the center of the gauge group, $Z_{N}$. Configurations of these $Z_{N}$ variables can be exactly mapped onto configurations of closed self-avoiding  surfaces, correspondingly, summation over $Z_{N}$ variables can be represented as a sum over all vortex configurations \cite{Irback:88}.

 For $Z_{N}$ lattice gauge theories there exists a duality transformation \cite{Ukawa:80} which allows to represent the observables either in terms of center vortices ($Z_{N}$ magnetic strings) or in terms of surfaces which open on Wilson loops. Since Wilson loop represents the worldline of an electric charge, it is reasonable to call such surfaces $Z_{N}$ electric strings. The aim of this paper is to investigate the representation of $SU\lr{2}$ lattice gauge theory in terms of a sum over all surfaces of such $Z_{2}$ electric strings and to compare the properties of electric strings and center vortices. It turns out that in contrast to center vortices, electric $Z_{2}$ strings are gauge-independent, thus the area of such strings can not be minimized by a procedure similar to the maximal center projection, which ensures the physical scaling of the density of center vortices \cite{DelDebbio:98, Polikarpov:03:1}. The density of $Z_{2}$ electric strings diverges in the continuum limit, consequently, they can not be described in terms of continuum theory. Nevertheless, there is a mechanism which makes the contribution of the minimal surface dominant -- namely, the statistical weight of electric strings in the partition function is not positively defined due to the presence of topological $Z_{2}$ monopoles.

 In order to separate $Z_{2}$ center variables in $SU\lr{2}$ lattice gauge theory, each lattice link variable should be represented as $g_{l} = \lr{-1}^{m_{l}} \tilde{g}_{l}$, $m_{l} = 0,1$. By definition the variables $\tilde{g}_{l}$ are the elements of the coset manifold $SU\lr{2}/Z_{2} = SO\lr{3}$. It will be assumed that multiplication of all "tilded" variables is a multiplication in $SO\lr{3}$ group. Note, however, that the product of two $SU\lr{2}$ group elements can not be in general expressed as the product of $\lr{-1}^{m_{l}}$ and the product of $\tilde{g}_{l}$ in $SO\lr{3}$. In order to characterize this deviation, it is useful to define the $Z_{2}$-valued function $m \lr{\tilde{g}_{1}, \tilde{g}_{2}}$ as:
\begin{eqnarray}
\label{SUNProduct}
g_{1} g_{2} = \lr{-1}^{m\lr{\tilde{g}_{1}, \tilde{g}_{2}} + m_{1} + m_{2}} \tilde{g}_{1} \tilde{g}_{2}
\end{eqnarray}
where $g_{1,2} = \lr{-1}^{m_{1,2}} \tilde{g}_{1,2}$. Such function can easily be generalized for any number of $SU\lr{2}$ variables. For the purposes of this paper it is convenient to define the $Z_{2}$-valued plaquette function $m_{p}$ and the functional $m\lrs{C}$ which characterize the difference of products of $SU\lr{2}$ and $SO\lr{3}$ link variables over lattice plaquettes and over arbitrary closed loops, respectively:
\begin{eqnarray}
\label{PlaquetteVar}
\prod \limits_{l \in p} g_{l} = \lr{-1}^{m_{p} + \sum \limits_{l \in p} m_{l}} \prod \limits_{l \in p} \tilde{g}_{l}
\nonumber \\
\prod \limits_{l \in C} g_{l} = \lr{-1}^{m\lrs{C} + \sum \limits_{l \in C} m_{l}} \prod \limits_{l \in C} \tilde{g}_{l}
\end{eqnarray}
By definition $m_{p}$ and $m\lrs{C}$ depend on the $SO\lr{3}$ variables $\tilde{g}_{l}$ only. The values of $m\lr{\tilde{g}_{1}, \tilde{g}_{2}}$, $m_{p}$ and $m\lrs{C}$ are in one-to-one correspondence with the homotopy classes $\pi_{1}\lr{SU\lr{2}/Z_{2}} \simeq Z_{2}$ \cite{Goddard:77:1}. In order to make this statement more precise, consider, for instance, the function $m\lr{\tilde{g}_{1}, \tilde{g}_{2}}$. The points $\tilde{g}_{1} \in SO\lr{3}$ and $\tilde{g}_{1} \tilde{g}_{2} \in SO\lr{3}$ can be connected by two different geodesics $\gamma_{1}$ and $\gamma_{2}$ in the following way:
\begin{eqnarray}
\label{HomotopyDef}
\gamma_{1}: \quad \tilde{g}\lr{s} = \tilde{g}_{1} \tilde{g}_{2}^{s}, \quad s \in \lrs{0, 1} \\
\gamma_{2}: \quad \tilde{g}\lr{s} = \tilde{g}_{1}^{1 - s} \tilde{g}_{2}, \quad s \in \lrs{0, 1}
\end{eqnarray}
The geodesics $\gamma_{1}$ and $\gamma_{2}$ form a loop on the group manifold, which is characterized by some element of $\pi_{1} \lr{ SO\lr{3} } \simeq Z_{2}$ which is precisely $\lr{-1}^{ m\lr{\tilde{g}_{1}, \tilde{g}_{2}} }$.

 Further analysis is most easily performed using the notations of external calculus on the lattice \cite{Becher:82:1}. $p$-forms on the lattice are associated with $p$-simplices, e.g. scalar functions are associated with lattice sites, $1$-forms - with lattice links, $2$-forms - with lattice plaquettes etc. Correspondingly, scalar functions will be denoted by a subscript $s$, $1$-forms -- by a subscript $l$ and $2$-forms -- by a subscript $p$. External and co-external differentials are denoted as $d$ and $\delta$, respectively. Scalar product of two $p$-forms $f$ and $g$ is denoted as $\lr{f,g}$ and the Hodge operator is denoted as $\hodge$. Hodge operator on the lattice acts between $p$-forms defined on the simplices of the original lattice and $D-p$-forms defined on the simplices of the dual lattice. It can be shown that the operators $d$, $\delta$, $\hodge$ and the scalar product of $p$-forms on the lattice have the same properties as in the continuum theory \cite{Becher:82:1}. In all operations on $Z_{2}$-valued forms summation is understood as summation modulo $2$.

 Note that gauge transformations in $SO\lr{3}$ gauge theory $\tilde{g}_{l} \rightarrow \tilde{h}_{s} \tilde{g}_{l} \tilde{h}_{s'} $ affect $m_{p}$ and $m \lrs{C}$: $m_{p} \rightarrow m_{p} + d \tilde{m}_{l} $, $m\lrs{C} \rightarrow m\lrs{C} + \sum \limits_{l \in C} \tilde{m}_{l}$, where $\tilde{m}_{l} = m\lr{\tilde{h}_{s},\tilde{g}_{l}} + m\lr{ \tilde{h}_{s} \tilde{g}_{l}, \tilde{h}^{-1}_{s'}}$. A gauge-invariant conserved current of topological $Z_{2}$ monopoles can be defined as follows \cite{Tomboulis:81:1}:
\begin{eqnarray}
\label{MonopoleCurrent}
j_{l\hodge} = \hodge d m_{p} = \delta \hodge m_{p}, \quad \delta j_{l \hodge} = 0
\end{eqnarray}

 The Wilson loop in the fundamental representation of $SU\lr{2}$ gauge group can be expressed in terms of the new variables $\tilde{g}_{l}$ and $m_{l}$ as:
\begin{eqnarray}
\label{WilsonLoopSO3Z21}
Z\lr{\beta} W\lrs{C}
 = \nonumber \\ =
\int \limits_{SU\lr{2}} \prod \limits_{l} dg_{l}
\tr\lr{\prod \limits_{l \in C} g_{l}}
\prod \limits_{p}
\expa{ \beta/2 \; \tr g_{p}}
 = \nonumber \\ =
\sum \limits_{m_{l}} \int \limits_{SO\lr{3}} \prod \limits_{l} d \tilde{g}_{l}
\tr \lr{\prod \limits_{l \in C} \tilde{g}_{l}} \:
\lr{-1}^{\sum \limits_{l \in C} m_{l} + m\lrs{C}}
\nonumber \\
\prod \limits_{p} \expa{ \beta/2 \lr{-1}^{d m_{l} + m_{p}} \tr \tilde{g}_{p} }
\end{eqnarray}
where $Z\lr{\beta}$ is the partition function of the theory, $g_{p} = \prod \limits_{l \in p} g_{l}$ and $\tilde{g}_{p} = \prod \limits_{l \in p} \tilde{g}_{l}$.

 The sum over $Z_{2}$-valued variables $m_{l}$ in (\ref{WilsonLoopSO3Z21}) can be represented as a sum over self-avoiding surfaces in two different ways, using the original or the dual lattice. One of these representations converges in the weak coupling limit, while the other is more suitable for the strong-coupling expansion \cite{Ukawa:80}. Configurations of center vortices ($Z_{2}$ magnetic strings) can be directly constructed from $m_{l}$'s. Namely, the worldsheets of center vortices are the surfaces on the dual lattice which consist of plaquettes $p \hodge$ for which $\hodge d m_{l} = 1$ \cite{Polikarpov:03:1, DelDebbio:98}. It is straightforward to show that such surfaces are indeed closed if one notes that for $Z_{2}$ - valued $2$-forms co-external derivative yields the $1$-form which is equal to $1$ only if the link belongs to the boundary of the surface made of plaquettes for which this $2$-form is nonzero. Since $\delta \hodge d m_{l} = \hodge d \hodge \hodge d m_{l} = \hodge d d m_{l} = 0$, the surfaces constructed from dual plaquettes with nonzero $\hodge d m_{l}$ are always closed. Such one-to-one correspondence between co-closed $Z_{2}$-valued $2$-forms and closed surfaces allows one to rewrite the sum over $m_{l}$'s as a sum over all closed non-intersecting vortex worldsheets $\Sigma_{m}$ on the dual lattice:
\begin{widetext}
\begin{eqnarray}
\label{WilsonLoopRSR_cv}
Z\lr{\beta} W\lrs{C} =
\sum \limits_{m_{l}} \int \limits_{SO\lr{3}} \prod \limits_{l} d \tilde{g}_{l}
\tr \lr{\prod \limits_{l \in C} \tilde{g}_{l}} \:
\lr{-1}^{\sum \limits_{l \in C} m_{l} + m\lrs{C}}
\nonumber \\
\prod \limits_{p} \expa{ \beta/2 \; \lr{-1}^{d m_{l} + m_{p}} \tr\tilde{g}_{p} }
 =
\sum \limits_{m_{p\hodge}; \delta m_{p\hodge} = 0}
\int \limits_{SO\lr{3}} \prod \limits_{l} d \tilde{g}_{l}
\tr \lr{\prod \limits_{l \in C} \tilde{g}_{l}} \:
\nonumber \\
\lr{-1}^{m\lrs{C} + \sum \limits_{p \in \Sigma_{C}} \hodge m_{p \hodge} }
\prod \limits_{p} \expa{ \beta/2 \; \lr{-1}^{m_{p\hodge} + m_{p}} \tr\tilde{g}_{p} }
 = \nonumber \\ =
\sum \limits_{\Sigma_{m}; \; \partial \Sigma_{m} = 0}
\lr{-1}^{ L\lrs{\Sigma_{m},C} }
\int \limits_{SO\lr{3}} \prod \limits_{l} d \tilde{g}_{l}
\tr \lr{\prod \limits_{l \in C} \tilde{g}_{l}} \:
\nonumber \\
\lr{-1}^{m\lrs{C}}
\prod \limits_{p} \expa{ \beta/2 \; \lr{-1}^{m_{p}} \tr\tilde{g}_{p} }
\prod \limits_{\hodge p \in \Sigma_{m}}
\expa{ - \beta \; \lr{-1}^{m_{p}} \tr\tilde{g}_{p} }
\end{eqnarray}
where $m_{p\hodge} = \hodge d m_{l}$, $\Sigma_{C}$ is an arbitrary surface spanned on the Wilson loop and $L\lrs{\Sigma_{m},C} = \sum \limits_{p \in \Sigma_{C}} \hodge m_{p \hodge}$ is the topological winding number of the surface $\Sigma_{m}$ and the loop $C$. The factor $L\lrs{\Sigma_{m},C}$ in (\ref{WilsonLoopRSR_cv}) implies that the Wilson loop changes sign each time it is crossed by center vortex. Note that the representation (\ref{WilsonLoopRSR_cv}) is gauge-dependent, since the action associated center vortices depends on the non-invariant functions $m_{p}$. The factor $\lr{-1}^{m\lrs{C}}$ is also gauge-dependent. Although the Wilson loop remains gauge-independent in any case, one can try to fix the gauge in such a way that the contribution of center vortices to the expectation value (\ref{WilsonLoopRSR_cv}) becomes dominant and the contribution of the terms $\lr{-1}^{m\lrs{C}}$ and $\tr \lr{\prod \limits_{l \in C} \tilde{g}_{l}}$ to the string tension can be neglected. It turns out that such gauge-fixing procedure is exactly the maximal center projection \cite{Polikarpov:03:1, DelDebbio:98, Bornyakov:01:1}, which rotates all link variables as close as possible to some element of the group center. It is also interesting to note that the presence of topological monopoles changes the action of center vortices, which indicates that monopoles can induce some nontrivial dynamics on the vortex worldsheet. Localization of Abelian monopoles on center vortices and the associated two-dimensional dynamics were indeed observed in lattice simulations \cite{Polikarpov:03:1, Buividovich:07:3}.

 The expression (\ref{WilsonLoopSO3Z21}) can be also represented as a sum over worldsheets of $Z_{2}$ electric strings -- closed self-avoiding surfaces $\Sigma_{e}$ which open on the loop $C$ \cite{Irback:88}. This is achieved by expanding the statistical weight of each plaquette in powers of $\lr{-1}^{d m_{l}}$ using the identity $\expa{\lr{-1}^{m} x} = \ch x \; \lr{1 + \lr{-1}^{m} \th{x}}$. The product of all the weights is then expanded into the sum of products of $\lr{-1}^{d m_{l}}$ over different sets of lattice plaquettes. After summation over $m_{l}$ each such product contributes to the sum only if each $m_{l}$ enters the product an even number of times. In this case the corresponding set of plaquettes forms a surface $\Sigma_{e}$ bounded by the loop $C$ \cite{Irback:88}, i.e. the sum over all sets of lattice plaquettes reduces to a sum over all such $\Sigma_{e}$:
\begin{eqnarray}
\label{WilsonLoopRSR}
Z\lr{\beta} W\lrs{C} =
\sum \limits_{\Sigma_{e}; \: \partial \Sigma_{e} = C}
\int \limits_{SO\lr{3}} \prod \limits_{l} d \tilde{g}_{l}
\tr \lr{\prod \limits_{l \in C} \tilde{g}_{l}} \;
\lr{-1}^{m\lrs{C} + \sum \limits_{\Sigma_{e \; min}} m_{p}}
\nonumber \\
\prod \limits_{p} \ch\lr{\beta/2 \; \tr\tilde{g}_{p}} \prod \limits_{p \in \Sigma_{e}} \th\lr{\beta/2 \; \tr\tilde{g}_{p}} \lr{-1}^{\sum \limits_{l\hodge \in V} j_{l\hodge}}
\end{eqnarray}
\end{widetext}
where $\Sigma_{e \; min}$ is the surface of the minimal area spanned on the loop $C$ and the Stokes theorem was used to represent $\sum \limits_{p \in \Sigma_{e}, \Sigma_{e \; min}} m_{p}$ as a sum over all links dual to cubes which belong to the volume $V$ bounded by $\Sigma_{e}$ and $\Sigma_{e \; min}$. Note that both combinations of $Z_{2}$ variables in (\ref{WilsonLoopRSR}), $m\lrs{C} + \sum \limits_{\Sigma_{e \; min}} m_{p}$ and $\sum \limits_{l\hodge \in V} j_{l\hodge}$, are gauge-invariant, therefore the representation (\ref{WilsonLoopRSR}) is gauge-independent. Thus the Wilson loop in the fundamental representation of $SU\lr{2}$ gauge group (or, more generally, in all representations with half-integer spin) can be represented as a sum over all surfaces of $Z_{2}$ electric strings $\Sigma_{e}$ which open on it. It is not difficult to derive similar representations for the partition function of the theory or another observables such as t'Hooft loop or the correlators of Wilson loops.

 In order to see whether the representations (\ref{WilsonLoopRSR}) and (\ref{WilsonLoopRSR_cv}) can be used in the continuum limit of the theory, one should study the scaling of the total area, or, equivalently, of the density of center vortices or $Z_{2}$ electric strings. Consider first the density of center vortices. According to (\ref{WilsonLoopRSR_cv}), some lattice plaquette $\hodge p$ belongs to center vortex if for the dual plaquette $p$ $\prod \limits_{l \in p} \sign \tr g_{l} = -1$, correspondingly, the probability that a given plaquette belongs to center vortex is:
\begin{eqnarray}
\label{cv_plaq_prob}
P = \langle \; \frac{1 - \prod \limits_{l \in p} \sign \tr g_{l}}{2} \; \rangle = \frac{1 - \langle \; \prod \limits_{l \in p} \sign \tr g_{l} \; \rangle }{2}
\end{eqnarray}
$P$ is nothing but the density of vortices in lattice units, which is by definition gauge-dependent. Consider first the theory without gauge-fixing and integrate $\prod \limits_{l \in p} \sign \tr g_{l}$ over gauge orbits $g_{l} \rightarrow h_{s} g_{l} h_{s'}^{-1}$. A simple calculation which involves character decomposition of $\sign \tr g$ and using the "gluing" formula for the integrals of group characters gives the result $\int \prod \limits_{s \in p} dh_{s} \prod \limits_{l \in p} \sign \tr\lr{ h_{s} g_{l} h_{s'}^{-1} } = \sum \limits_{k = 1/2, 3/2, \ldots} \alpha_{k} \chi_{k}\lr{g_{p}}$, where the coefficients $\alpha_{k}$ behave as $ \lr{-1}^{k+1/2} k^{-6}$. Thus integrating (\ref{cv_plaq_prob}) over gauge orbits yields some gauge-invariant function of the plaquette variable $g_{p}$. It can be shown that this function takes values in the range $\lrs{ -1/3 + 2/\pi^{2}, 1/3 - 2/\pi^{2}}$. As $1/3 - 2/\pi^{2} \approx 0.130691 < 1$, the density of center vortices in lattice units remains finite at weak coupling, and their physical density diverges as $a^{-4}$. On the other hand, any perturbative calculation around the vacuum $g_{l}=1$ yields exactly zero density of center vortices, which thus appear as truly nonperturbative objects. The situation changes dramatically after the maximal center gauge is imposed -- in this case the total area of center vortices is minimized and their density in lattice units goes to zero as $a^{2}$, thus their physical density remains finite in the continuum limit \cite{DelDebbio:98, Polikarpov:03:1}.

 Since statistical weights of $Z_{2}$ electric strings in the sum (\ref{WilsonLoopRSR}) are not positive, strictly speaking they can not be interpreted as random surfaces \cite{Ambjorn:94:1}. Nevertheless, one can define the vacuum expectation value of their density by introducing the "chemical potential" $\mu_{e}$ for electric strings and by differentiating the partition function $Z\lr{\beta, \mu_{e}}$ over it. This amounts to multiplying the statistical weight of each surface by an additional factor $\expa{ - \mu_{e} |\Sigma_{e}| }$, where $|\Sigma_{e}|$ is the total area of $\Sigma_{e}$. The partition function of the theory can be calculated from (\ref{WilsonLoopRSR}) by shrinking the loop $C$ to zero. After such modification of (\ref{WilsonLoopRSR}) one can reverse all the transformations which led to this representation and write the partition function of the theory at nonzero $\mu_{e}$ in terms of original link variables as:
\begin{eqnarray}
\label{ElectricStringsPF_mu}
Z\lr{\beta, \mu_{e}} = \int \limits_{SU\lr{2}} dg_{l}
\nonumber \\
\prod \limits_{p} \lr{\ch\lr{\beta/2 \; \tr g_{p}} + e^{-\mu_{e}} \sh\lr{\beta/2 \; \tr g_{p}}}
\end{eqnarray}
This partition function interpolates between the partition functions of $SU\lr{2}$ and $SO\lr{3}$ lattice gauge theories at $\mu_{e}=0$ and $\mu_{e} \rightarrow \infty$ respectively. 

 Now the average total area of $\Sigma_{e}$ in lattice units can be found by differentiating (\ref{ElectricStringsPF_mu}) over $\mu_{e}$:
\begin{eqnarray}
\label{ElectricStringsArea}
\langle \: |\Sigma_{e}| \: \rangle = - \frac{\partial}{\partial \mu_{e}} \: \ln Z\lr{\beta, \mu_{e}}|_{\mu_{e} = 0}
 = \nonumber \\ =
\sum \limits_{p} \frac{1 - \langle \: \expa{ - \beta \tr g_{p} } \: \rangle}{2}
\end{eqnarray}
The expectation value $\langle \: \expa{ - \beta \tr g_{p} } \: \rangle$ tends to zero as $\beta \rightarrow \infty$ and the continuum limit is approached, therefore in the continuum limit of the theory $Z_{2}$ electric strings occupy half of all lattice plaquettes and their physical density diverges as $a^{-4}$. Nevertheless, the sum (\ref{WilsonLoopRSR}) remains well-defined at $a \rightarrow 0$ and in fact sums up to $\expa{ - \sigma |\Sigma_{e \; min}| }$, which can only be explained by the exact cancellation of contributions from different surfaces with opposite signs of $\lr{-1}^{\sum \limits_{l\hodge \in V} j_{l\hodge}}$, i.e. due to the presence of $Z_{2}$ topological monopoles. Indeed, if the term with $j_{l \hodge}$ is omitted, the expression (\ref{WilsonLoopRSR}) can be considered as the partition function of $Z_{2}$ lattice gauge theory with fluctuating, but always positive coupling. It can be shown that in the weak coupling limit electric strings in $Z_{2}$ gauge theory also occupy half of all lattice plaquettes. The sum over such creased surfaces with positive weights can only lead to perimeter dependence of the Wilson loop, which is indeed the case for $Z_{N}$ lattice gauge theories in the weak coupling limit \cite{Ukawa:80}. The terms $\tr \lr{\prod \limits_{l \in C} \tilde{g}_{l}}$ and $\lr{-1}^{m\lrs{C} + \sum \limits_{\Sigma_{e \; min}} m_{p}}$ are also not likely to contribute to the full string tension, since it can be shown that the expression (\ref{WilsonLoopRSR}) yields physical string tension even when these terms are omitted \cite{Greensite:98}. Thus it is reasonable to conjecture that in the weak coupling limit $Z_{2}$ electric strings are confining due to the presence of topological monopoles with currents $j_{l\hodge}$. It could be interesting to study numerically the properties of such topological monopoles.

 To conclude, it was shown that unlike $Z_{2}$ center vortices, which remain physical in the continuum limit \cite{DelDebbio:98, Polikarpov:03:1}, their duals -- $Z_{2}$ electric strings -- can not be consistently described as random surfaces in the continuum theory. Instead, electric strings condense in a creased phase with infinite Hausdorf dimension, but nevertheless due to some cancelations between surfaces with positive and negative statistical weights the minimal surface $\Sigma_{e \; min}$ dominates in the Wilson loop. In fact the formation of some creased structures is typical for  subcritical strings \cite{Ambjorn:94:1}. For instance, subcritical Nambu-Goto strings exist only as branched polymers \cite{Ambjorn:94:1}. It was conjectured in \cite{Polyakov:04:1} that such subcritical strings can be described as strings on $AdS_{5}$ background, which hints at some possible relation with AdS/QCD.

\begin{acknowledgments}
 This work was partly supported by grants RFBR 05-02-16306, 07-02-00237-a, by the EU Integrated Infrastructure Initiative Hadron Physics (I3HP) under contract RII3-CT-2004-506078, by Federal Program of the Russian Ministry of Industry, Science and Technology No 40.052.1.1.1112 and by Russian Federal Agency for Nuclear Power.
\end{acknowledgments}

%\bibliography{MyBibliography}

\begin{thebibliography}{16}
\expandafter\ifx\csname natexlab\endcsname\relax\def\natexlab#1{#1}\fi
\expandafter\ifx\csname bibnamefont\endcsname\relax
  \def\bibnamefont#1{#1}\fi
\expandafter\ifx\csname bibfnamefont\endcsname\relax
  \def\bibfnamefont#1{#1}\fi
\expandafter\ifx\csname citenamefont\endcsname\relax
  \def\citenamefont#1{#1}\fi
\expandafter\ifx\csname url\endcsname\relax
  \def\url#1{\texttt{#1}}\fi
\expandafter\ifx\csname urlprefix\endcsname\relax\def\urlprefix{URL }\fi
\providecommand{\bibinfo}[2]{#2}
\providecommand{\eprint}[2][]{\url{#2}}

\bibitem[{\citenamefont{Polyakov}(2004)}]{Polyakov:04:1}
\bibinfo{author}{\bibfnamefont{A.~M.} \bibnamefont{Polyakov}},
  \emph{\bibinfo{title}{Confinement and liberation}} (\bibinfo{year}{2004}),
  \urlprefix\url{http://arxiv.org/abs/hep-th/0407209}.

\bibitem[{\citenamefont{Karch et~al.}(2006)\citenamefont{Karch, Katz, Son, and
  Stephanov}}]{Son:06:1}
\bibinfo{author}{\bibfnamefont{A.}~\bibnamefont{Karch}},
  \bibinfo{author}{\bibfnamefont{E.}~\bibnamefont{Katz}},
  \bibinfo{author}{\bibfnamefont{D.~T.} \bibnamefont{Son}}, \bibnamefont{and}
  \bibinfo{author}{\bibfnamefont{M.}~\bibnamefont{Stephanov}},
  \bibinfo{journal}{Physical Review D}  (\bibinfo{year}{2006}),
  \urlprefix\url{http://arxiv.org/abs/hep-ph/0602229}.

\bibitem[{\citenamefont{Bali et~al.}(1995)\citenamefont{Bali, Schilling, and
  Schlichter}}]{Bali:95:1}
\bibinfo{author}{\bibfnamefont{G.~S.} \bibnamefont{Bali}},
  \bibinfo{author}{\bibfnamefont{K.}~\bibnamefont{Schilling}},
  \bibnamefont{and}
  \bibinfo{author}{\bibfnamefont{C.}~\bibnamefont{Schlichter}},
  \bibinfo{journal}{Physical Review D} \textbf{\bibinfo{volume}{51}},
  \bibinfo{pages}{5165} (\bibinfo{year}{1995}),
  \urlprefix\url{http://arxiv.org/abs/hep-lat/9409005}.

\bibitem[{\citenamefont{Boyko et~al.}(2007)\citenamefont{Boyko, Gubarev, and
  Morozov}}]{Gubarev:07:1}
\bibinfo{author}{\bibfnamefont{P.~Y.} \bibnamefont{Boyko}},
  \bibinfo{author}{\bibfnamefont{F.~V.} \bibnamefont{Gubarev}},
  \bibnamefont{and} \bibinfo{author}{\bibfnamefont{S.~M.}
  \bibnamefont{Morozov}}, \emph{\bibinfo{title}{On the fine structure of {QCD}
  confining string}} (\bibinfo{year}{2007}),
  \urlprefix\url{http://arxiv.org/abs/0704.1203}.

\bibitem[{\citenamefont{{t' Hooft}}(1978)}]{tHooft:78}
\bibinfo{author}{\bibfnamefont{G.}~\bibnamefont{{t' Hooft}}},
  \bibinfo{journal}{Nuclear Physics B} \textbf{\bibinfo{volume}{138}},
  \bibinfo{pages}{1 } (\bibinfo{year}{1978}).

\bibitem[{\citenamefont{{Del Debbio} et~al.}(1998)\citenamefont{{Del Debbio},
  Faber, Giedt, Greensite, and Olejnik}}]{DelDebbio:98}
\bibinfo{author}{\bibfnamefont{L.}~\bibnamefont{{Del Debbio}}},
  \bibinfo{author}{\bibfnamefont{M.}~\bibnamefont{Faber}},
  \bibinfo{author}{\bibfnamefont{J.}~\bibnamefont{Giedt}},
  \bibinfo{author}{\bibfnamefont{J.}~\bibnamefont{Greensite}},
  \bibnamefont{and} \bibinfo{author}{\bibfnamefont{S.}~\bibnamefont{Olejnik}},
  \bibinfo{journal}{Physical Review D} \textbf{\bibinfo{volume}{58}},
  \bibinfo{pages}{094501} (\bibinfo{year}{1998}),
  \urlprefix\url{http://arxiv.org/abs/hep-lat/9801027}.

\bibitem[{\citenamefont{Gubarev et~al.}(2003)\citenamefont{Gubarev, Kovalenko,
  Polikarpov, Syritsyn, and Zakharov}}]{Polikarpov:03:1}
\bibinfo{author}{\bibfnamefont{F.~V.} \bibnamefont{Gubarev}},
  \bibinfo{author}{\bibfnamefont{A.~V.} \bibnamefont{Kovalenko}},
  \bibinfo{author}{\bibfnamefont{M.~I.} \bibnamefont{Polikarpov}},
  \bibinfo{author}{\bibfnamefont{S.~N.} \bibnamefont{Syritsyn}},
  \bibnamefont{and} \bibinfo{author}{\bibfnamefont{V.~I.}
  \bibnamefont{Zakharov}}, \bibinfo{journal}{Physics Letters B}
  \textbf{\bibinfo{volume}{574}}, \bibinfo{pages}{136 } (\bibinfo{year}{2003}),
  \urlprefix\url{http://arxiv.org/abs/hep-lat/0212003}.

\bibitem[{\citenamefont{Buividovich and Polikarpov}(2007)}]{Buividovich:07:3}
\bibinfo{author}{\bibfnamefont{P.~V.} \bibnamefont{Buividovich}}
  \bibnamefont{and} \bibinfo{author}{\bibfnamefont{M.~I.}
  \bibnamefont{Polikarpov}}, \bibinfo{journal}{Nuclear Physics B}
  \textbf{\bibinfo{volume}{786}}, \bibinfo{pages}{84 } (\bibinfo{year}{2007}),
  \urlprefix\url{http://arxiv.org/abs/0705.3745}.

\bibitem[{\citenamefont{Irb{\"{a}}ck}(1988)}]{Irback:88}
\bibinfo{author}{\bibfnamefont{A.}~\bibnamefont{Irb{\"{a}}ck}},
  \bibinfo{journal}{Physics Letters B} \textbf{\bibinfo{volume}{211}},
  \bibinfo{pages}{129 } (\bibinfo{year}{1988}).

\bibitem[{\citenamefont{Ukawa et~al.}(1980)\citenamefont{Ukawa, Windey, and
  Guth}}]{Ukawa:80}
\bibinfo{author}{\bibfnamefont{A.}~\bibnamefont{Ukawa}},
  \bibinfo{author}{\bibfnamefont{P.}~\bibnamefont{Windey}}, \bibnamefont{and}
  \bibinfo{author}{\bibfnamefont{A.~H.} \bibnamefont{Guth}},
  \bibinfo{journal}{Physical Review D} \textbf{\bibinfo{volume}{21}},
  \bibinfo{pages}{1013 } (\bibinfo{year}{1980}),
  \urlprefix\url{http://prola.aps.org/abstract/PRD/v21/i4/p1013_1}.

\bibitem[{\citenamefont{Goddard et~al.}(1977)\citenamefont{Goddard, Nuyts, and
  Olive}}]{Goddard:77:1}
\bibinfo{author}{\bibfnamefont{P.}~\bibnamefont{Goddard}},
  \bibinfo{author}{\bibfnamefont{J.}~\bibnamefont{Nuyts}}, \bibnamefont{and}
  \bibinfo{author}{\bibfnamefont{D.}~\bibnamefont{Olive}},
  \bibinfo{journal}{Nuclear Physics B} \textbf{\bibinfo{volume}{125}},
  \bibinfo{pages}{1 } (\bibinfo{year}{1977}).

\bibitem[{\citenamefont{Becher and Joos}(1982)}]{Becher:82:1}
\bibinfo{author}{\bibfnamefont{P.}~\bibnamefont{Becher}} \bibnamefont{and}
  \bibinfo{author}{\bibfnamefont{H.}~\bibnamefont{Joos}}, \bibinfo{journal}{Z.
  Phys. C} \textbf{\bibinfo{volume}{15}}, \bibinfo{pages}{343}
  (\bibinfo{year}{1982}).

\bibitem[{\citenamefont{Tomboulis}(1981)}]{Tomboulis:81:1}
\bibinfo{author}{\bibfnamefont{E.~T.} \bibnamefont{Tomboulis}},
  \bibinfo{journal}{Physical Review D} \textbf{\bibinfo{volume}{23}},
  \bibinfo{pages}{2371 } (\bibinfo{year}{1981}),
  \urlprefix\url{http://prola.aps.org/abstract/PRD/v23/i10/p2371_1}.

\bibitem[{\citenamefont{Bornyakov et~al.}(2001)\citenamefont{Bornyakov,
  Komarov, and Polikarpov}}]{Bornyakov:01:1}
\bibinfo{author}{\bibfnamefont{V.~G.} \bibnamefont{Bornyakov}},
  \bibinfo{author}{\bibfnamefont{D.~A.} \bibnamefont{Komarov}},
  \bibnamefont{and} \bibinfo{author}{\bibfnamefont{M.~I.}
  \bibnamefont{Polikarpov}}, \bibinfo{journal}{Physics Letters B}
  \textbf{\bibinfo{volume}{497}}, \bibinfo{pages}{151} (\bibinfo{year}{2001}),
  \urlprefix\url{http://arxiv.org/abs/hep-lat/0009035}.

\bibitem[{\citenamefont{Ambj{\o}rn}(1994)}]{Ambjorn:94:1}
\bibinfo{author}{\bibfnamefont{J.}~\bibnamefont{Ambj{\o}rn}},
  \emph{\bibinfo{title}{Quantization of geometry}},
  \bibinfo{howpublished}{Lectures presented at the 1994 Les Houches Summer
  School} (\bibinfo{year}{1994}),
  \urlprefix\url{http://arxiv.org/abs/hep-th/9411179}.

\bibitem[{\citenamefont{Greensite et~al.}(1999)\citenamefont{Greensite, Faber,
  and Olejnik}}]{Greensite:98}
\bibinfo{author}{\bibfnamefont{J.}~\bibnamefont{Greensite}},
  \bibinfo{author}{\bibfnamefont{M.}~\bibnamefont{Faber}}, \bibnamefont{and}
  \bibinfo{author}{\bibfnamefont{S.}~\bibnamefont{Olejnik}},
  \bibinfo{journal}{JHEP} \textbf{\bibinfo{volume}{9901}}, \bibinfo{pages}{008}
  (\bibinfo{year}{1999}), \urlprefix\url{http://arxiv.org/abs/hep-lat/9810008}.

\end{thebibliography}
%\bibliographystyle{apsrev}

\fussy

\end{document}